\documentclass[12pt]{article}

\usepackage{epsfig}
\usepackage{graphicx}
\usepackage{natbib}
\usepackage[usenames,dvipsnames]{color}
\usepackage{amssymb}
\usepackage{amsmath} 
\usepackage{mathrsfs}
\usepackage[colorlinks,citecolor=blue,urlcolor=blue]{hyperref}
\usepackage{bm}

\usepackage{setspace}

\usepackage[font = {it}]{caption}

\usepackage{verbatim}
\usepackage{lineno}
\usepackage{times}
\usepackage{soul}
\usepackage{color}
\usepackage{enumerate}
\usepackage{times}
\usepackage{changepage}
\usepackage{multirow}
\usepackage{booktabs}
\usepackage{pdflscape}
\usepackage{hyperref}

\newcommand{\alphahat}{\hat{\alpha}}

\newcommand{\bB}{\ensuremath{\mathbf{B}}}

\newcommand{\bk}{\ensuremath{\mathbf{k}}}

\newcommand{\bs}{\ensuremath{\mathbf{s}}}

\newcommand{\iid}{\stackrel{\mathrm{iid}}{\sim}}

\newcommand{\dd}{\; \text{d} }

\newcommand{\calS}{{\cal S}}

\newcommand{\beq}{ \begin{equation}}
\newcommand{\eeq}{ \end{equation}}
\newcommand{\beqn}{ \begin{eqnarray}}
\newcommand{\eeqn}{ \end{eqnarray}}

\newcommand{\eref}[1]{(\ref{#1})}
\newcommand{\fref}[1]{Figure~\ref{#1}}

\newcommand{\sref}[1]{Section~\ref{#1}}
\newcommand{\aref}[1]{Appendix~\ref{#1}}
\newcommand{\cref}[1]{Chapter~\ref{#1}}

\newcommand*\patchAmsMathEnvironmentForLineno[1]{%
  \expandafter\let\csname old#1\expandafter\endcsname\csname #1\endcsname
  \expandafter\let\csname oldend#1\expandafter\endcsname\csname end#1\endcsname
  \renewenvironment{#1}%
     {\linenomath\csname old#1\endcsname}%
     {\csname oldend#1\endcsname\endlinenomath}}%
\newcommand*\patchBothAmsMathEnvironmentsForLineno[1]{%
  \patchAmsMathEnvironmentForLineno{#1}%
  \patchAmsMathEnvironmentForLineno{#1*}}%
\AtBeginDocument{%
\patchBothAmsMathEnvironmentsForLineno{equation}%
\patchBothAmsMathEnvironmentsForLineno{align}%
\patchBothAmsMathEnvironmentsForLineno{flalign}%
\patchBothAmsMathEnvironmentsForLineno{alignat}%
\patchBothAmsMathEnvironmentsForLineno{gather}%
\patchBothAmsMathEnvironmentsForLineno{multline}%
}

\begin{document}
\thispagestyle{empty}
\baselineskip=28pt
\vskip 5mm
\begin{center} {\Large{\bf Exploration and inference in spatial extremes using empirical basis functions}}
\end{center}

\baselineskip=12pt

\vskip 5mm

\begin{center}
\large
Samuel A. Morris$^{1}$, Brian J. Reich$^1$ and Emeric Thibaud$^2$
\end{center}

\footnotetext[1]{
\baselineskip=10pt North Carolina State University}
\footnotetext[2]{
\baselineskip=10pt  Ecole Polytechnique F\'ed\'erale de Lausanne}

\baselineskip=17pt
\vskip 4mm
\vskip 6mm

\begin{center}
{\large{\bf Abstract}}
\end{center}

Statistical methods for inference on spatial extremes of large datasets are yet to be developed. Motivated by standard dimension reduction techniques used in spatial statistics, we propose an approach based on empirical basis functions to explore and model spatial extremal dependence. Based on a low-rank max-stable model we propose a data-driven approach to estimate meaningful basis functions using empirical pairwise extremal coefficients. These spatial empirical basis functions can be used to visualize the main trends in extremal dependence. In addition to exploratory analysis, we describe how these functions can be used in a Bayesian hierarchical model to model spatial extremes of large datasets. We illustrate our methods on extreme precipitations in eastern U.S.
\baselineskip=16pt


\pagenumbering{arabic}
\baselineskip=24pt


\newpage

\section{Introduction}\label{ebs:intro}

The spatial extreme value analysis (EVA) literature is expanding rapidly \citep[see, e.g.,][]{Davison2012,Davison2013,Dey.Yan:2015} to meet the demands of researchers to improve estimates of rare-event probabilities by borrowing information across space and to estimate the probability of extreme events occurring simultaneously at multiple locations.
Environmental datasets commonly include observations from hundreds or thousands of locations, and advanced tools are required to explore and analyze these data. 
For Gaussian data, Principle Components Analysis \citep[PCA,][]{Everitt2008}, also known as Empirically Orthogonal Functions \citep[EOFs,][]{Toggweiler2001}, has proven to be a powerful tool to study correlation between spatial locations; understand the most important large-scale spatial features; and reduce the dimension of the problem to allow for simple computation even for massive datasets.
Computation and exploration are arguably more difficult for EVA than Gaussian data, yet to our knowledge no tool analogous to spatial PCA has been developed for EVA. \citet{Bernard2013} proposed a clustering method for spatial extremes, which can be interpreted as a dimension reduction method; our approach is different and is motivated by identifying drivers of spatial extremes much like PCA and EOF approaches do. 

In EVA, extremes are separated from the bulk of the distribution by either analyzing only points above a threshold or block maxima \citep{Coles2001}, e.g., the daily precipitation values exceeding a high threshold or their annual maxima.
A natural spatial model for block maxima at several spatial locations is the max-stable process, which, under certain conditions, arises as the limit of the location-wise linearly scaled pointwise maxima of infinitely-many spatial processes \citep[Chapter~9]{deHaan2006}.
Max-stable processes can also be used to model spatial exceedances over a high threshold \citep{Thibaud2013,Huser2014}.

Inference for max-stable models is challenging because their likelihoods are intractable in large dimensions. \citet{Padoan2010} proposed the use of composite likelihoods for estimating parameters of max-stable models. More efficient approaches based on full likelihoods were recently developed for extremal threshold exceedances models \citep{Wadsworth2014,Engelke2015,Thibaud2013a} but these do not apply for max-stable models fitted to block maxima. Alternatively, non-max-stable models that retain extremal dependence such as the skew-$t$ process of \citet{Morris2016} can be used for extremes. \citet{Thibaud2015} showed how fully-Bayesian analysis can be performed for max-stable models, but the approach is cumbersome for large data sets. \citet{Reich2012} proposed a Bayesian max-stable model, based on a low-rank method based on spatial kernel functions. Although the previous model can be fit to large datasets, the Bayesian inference is computationally intensive. In this paper we will build on the previous model and show how the inference can be improved by considering a data-based low-rank approximation of the max-stable process.

The spectral representation \citep[\S9.6]{deHaan2006} states that any max-stable process can be represented in terms of a countable number of spatial processes, for example using Gaussian processes \citep{Schlather2002} or log Gaussian processes \citep{Kabluchko2009}.
In this paper we propose an empirical basis function (EBF) approach that builds on a finite truncation of the spectral representation of max-stable process, as in \citet{Wang2011}, and develops a method-of-moments estimator for the underlying spatial processes.
Unlike PCA/EOFs, but similar to dictionary learning \citep{Mairal2014} and non-negative matrix factorizations \citep{Lee1999}, the EBFs are not orthogonal.
Nonetheless these spatial functions can be plotted for exploratory analysis to reveal important spatial trends. 
In addition to exploratory analysis, we describe how the EBFs can be used for Bayesian inference. 
By basing the spatial dependence on EBFs, the resulting spatial analysis does not require dubious assumptions such as stationarity.
In addition, a Bayesian analysis for either block-maxima or point above a threshold is computationally feasible for large datsets because the entire spatial process is represented by a small number of basis functions.

The paper proceeds as follows. In \sref{ebs:model} we present the low-rank model. \sref{ebs:estimation} describes the algorithm used to estimate the spatial basis functions, and \sref{ebs:MCMC} describes the use of EBFs in a Bayesian hierarchical model. 
In \sref{ebs:sims} we present the results of a simulation study to evaluate the performance of our method for estimating basis functions. 
In \sref{ebs:analysis} we demonstrate the use of the EBFs for an analysis of precipitation data in the eastern U.S. Lastly in \sref{ebs:con} we give some summary conclusions and a brief discussion of the findings.

\section{Low-rank max-stable model}\label{ebs:model} Let $Y_{t}(\bs)$ be the observation process at spatial location $\bs\in\calS$ and time $t$; $\calS$ is a compact set in $\mathbb{R}^2$.  We temporarily drop the subscript $t$ and describe the model for the process $Y(\bs)$ for a single time point, but return to the spatiotemporal notation in \sref{ebs:estimation}.

Spatial dependence is captured by modeling $Y(\bs)$ as a max-stable process.
Max-stable processes have generalized extreme value (GEV) marginal distributions. At each location $\bs\in\calS$ the GEV distribution is
\begin{equation}\label{ebeq:GEVmarg}
F_{\bs}(y) =  \Pr\{Y(\bs) \leq y \}  = \exp \left(- \left[1 + \xi(\bs) \left\{\dfrac{y - \mu(\bs)}{\sigma(\bs)}\right\}\right]_+^{-1/\xi(\bs)}\right),
\end{equation}
where $x_+=\max(0,x)$. This distribution is denoted GEV$\{\mu(\bs),\sigma(\bs),\xi(\bs)\}$ and has three real parameters: location $\mu(\bs)$, scale $\sigma(\bs)>0$ and shape $\xi(\bs)$; the case $\xi(\bs)=0$ is defined as a limit in~\eqref{ebeq:GEVmarg}.
Spatial dependence is present both in the GEV parameters and in the standardized residual process  $Z(\bs) =  - \left[ \log F_{\bs}\{Y(\bs)\} \right]^{-1}$ which is max-stable and has unit Fr\'echet, i.e., GEV(1,1,1), marginal distribution for all $\bs$.

Our objective is to identify a low-rank max-stable model for the spatial dependence of the residual process $Z(\bs)$.
The spectral representation discussed in \citet[Theorem~9.6.1]{deHaan2006} shows that any max-stable process $Z(\bs)$ with unit Fr\'echet margins and continuous sample paths can be written as
\begin{align}\label{ebeq:spectral}
  Z(\bs) = \max_{l = 1,2,\ldots} B(\bs, \bk_l)q_{l},
\end{align}
where the functions $B(\bs, \bk_l)$ are nonnegative and satisfy $\int B(\bs, \bk_l) \dd \bk_l = 1$ for all $\bs\in\calS$, and $\int \sup_{\bs\in\calS} B(\bs, \bk_l) \dd \bk_l < \infty$, and $(\bk_1, q_1),  (\bk_2, q_2), \ldots$ are the points of a Poisson process with intensity measure ${\rm d}\bk\times ({\rm d} q/q^2)$ on $\calS \times (0,\infty)$. Conversely, a max-stable process with unit Fr\'echet marginals is obtained by choosing particular functions $B(\bs, \bk_l)$ satisfying the conditions stated above \citep[Theorem~9.6.1]{deHaan2006}; this has been used to construct max-stable models \citep[e.g.,][]{Smith1990,Schlather2002,Kabluchko2009}. In several max-stable models, such as \citet{Smith1990} and \citet{Reich2012}, the $\bk_l$ are spatial locations that represent the center of process $B(\bs ,\bk_l)$; however, in our proposed method the basis functions are not associated with one particular location and so to simplify notation we let $B_l(\bs) = B(\bs, \bk_l)$.

To arrive at a low-rank model, we assume that there is a finite and known number $L$ of spatial basis functions $B_1(\bs), \ldots, B_L(\bs)$ that explain the important spatial variation in the process. We motivate our approach as follows. \citet{Smith1990} uses the physical analogy that in \eref{ebeq:spectral}, $\bk_l$ is storm $l$'s spatial center, $B(\bs, \bk_l)$ is the storm's spatial extent, and $q_l$ is the magnitude of its precipitation.  The response $Z(\bs)$ is thus the maximum of the precipitation over the year's storms at site $\bs$. Following this analogy, our dimension reduction approach identifies the $L$ most influential storm patterns, $B_1(\bs),\ldots,B_L(\bs)$, and expresses year-to-year variation in terms of the annual intensity of these $L$ storm patterns.

Assuming a finite number $L$ of basis function we can rearrange \eqref{ebeq:spectral} to write
\begin{align} \label{ebeq:spectraltrunc}
 Z(\bs) = \max_{l = 1,\ldots,L} B_l(\bs)Z_{l},
\end{align}
where $Z_1,\ldots,Z_L$ are unit Fr\'echet random variables and the nonnegative basis functions are restricted so that $\sum_{l = 1}^L B_l(\bs) = 1$ for all $\bs$. Model~\eqref{ebeq:spectraltrunc} is known as a max-linear process. It can be shown that any max-stable process can be arbitrary well approximated by a low-rank max-linear model for some suitable choice of basis functions. This approximation was used by \citet{Wang2011} to obtain approximate conditional simulations from max-stable models.

Because it is unrealistic to assume that extreme realizations are exactly functions of $L$ basis functions, the process \eqref{ebeq:spectraltrunc} is inappropriate for modeling complex data based on a small number $L$ of basis functions. Hence we use a different low-rank max-stable model proposed by \citet{Reich2012}, which is based on multivariate max-stable models from \citet{Fougeres.etal:2009} and \citet{Fougeres.etal:2013}, and which can be seen as a noisy version of model~\eqref{ebeq:spectraltrunc}. We consider the model
\begin{align}\label{ebeq:ReichShabymodel}
  Z(\bs) = \theta(\bs) \epsilon(\bs) \quad \mathrm{where} \quad \theta(\bs) = \left\{\sum_{l=1}^L B_{l}(\bs)^{1/\alpha}A_{l}\right\}^{\alpha},
\end{align}
and where $\varepsilon(\bs)\iid$ GEV$(1,\alpha,\alpha)$, $B_{l}(\bs)\geq0$, $\sum_{l=1}^LB_{l}(\bs)=1$ for all $\bs$, and the $A_{l}$ have positive stable (PS; \aref{eba:gridapprox}) distribution, \mbox{$A_{l}\iid$ PS$(\alpha)$}, $\alpha\in(0,1)$; these conditions ensure that $Z(\bs)$ is max-stable and has unit Fr\'echet marginal distributions. The functions $B_l(\bs)$ control the shape of spatial dependence in the process $Z(\bs)$ while the parameter $\alpha$ acts as a nugget effect introducing independent noise at each location $\bs$ and thus giving discontinuous realizations of $Z(\bs)$. \citet[Fig.~1]{Reich2012} shows the effects of the basis functions and $\alpha$ on realizations of $Z(\bs)$. As $\alpha\rightarrow1$ the model converges to independence, i.e., a process with independent unit Fr\'echet marginals.
As $\alpha\rightarrow 0$ multivariate distributions of model~\eqref{ebeq:ReichShabymodel} converge to those of the max-linear model~\eqref{ebeq:spectraltrunc} \citep[][Section~3]{Reich2012}.

\citet{Reich2012} used standardized Gaussian kernel functions (GKFs) as spatial basis functions in their max-stable model. For $\bk_1, \ldots, \bk_L$ a fixed collection of $L$ spatial knot locations in $\calS$, they defined
\begin{align}\label{eq:gkf}
B_l(\bs) = \dfrac{\exp\left\{- (\|\bs - \bk_l\|/\rho)^2 \right\}}{\sum_{j = 1}^L \exp\left\{- (\|\bs - \bk_j\|/ \rho)^2 \right\}},
\end{align}
where $\|\cdot\|$ is the Euclidean norm and $\rho>0$ is a bandwidth parameter that they estimated as part of their Bayesian model. For a large and dense grid of knots $\bk_1, \ldots, \bk_L$ the GKF model provides a good approximation to the max-stable model of \citet{Smith1990}. In this paper, rather than selecting a predefined form for the basis functions $B_l$ we will estimate them from the data.

Extremal spatial dependence for max-stable processes can be summarized by the pairwise extremal coefficient $\vartheta(\bs_1, \bs_2)\in[1,2]$ defined such that
\begin{equation*}
  \Pr\{Z(\bs_1)\leq z, Z(\bs_2) \leq z\} = \Pr\{Z(\bs_1) \leq z\}^{\vartheta(\bs_1,\bs_2)},
\end{equation*}
with $\vartheta(\bs_1,\bs_2)=1$ and $\vartheta(\bs_1,\bs_2)=2$ corresponding to perfect dependence and independence of $Z(\bs_1)$ and $Z(\bs_2)$ respectively.
For the positive stable random effects model \eqref{ebeq:ReichShabymodel} it can be shown that the extremal coefficient for $\bs_1\neq \bs_2$ is
\begin{align*}
   \vartheta(\bs_1,\bs_2) = \sum_{l=1}^L \left\{B_{l}(\bs_1)^{1/\alpha}+B_{l}(\bs_2)^{1/\alpha}\right\}^\alpha.
\end{align*}
In particular, $\lim_{\|\bs_1-\bs_2\| \rightarrow 0}\vartheta(\bs_1,\bs_2) = 2^{\alpha}$, showing how $\alpha$ can be interpreted as a nugget effect in the model, as creating a discontinuity in the extremal coefficient function; note that $\vartheta(\bs,\bs)=1$ by definition.

In the next sections we will explore and model residual dependence in the process $Y(\bs)$ using model \eqref{ebeq:ReichShabymodel}. The motivation for our approach is that model \eqref{ebeq:ReichShabymodel} can be used as an approximation to most max-stable processes for a sufficiently large number $L$ of basis functions $B_l(\bs)$ suitably chosen. The functions $B_l(\bs)$ represents the main trends in the max-stable process $Z(\bs)$. We propose to estimate these basis functions, so-called empirical basis functions (EBFs), using extremal coefficients and then show how inference for $Y(\bs)$ can be based on these EBFs.

\section{Estimating the basis functions}\label{ebs:estimation} We describe the estimation of the EBFs for model \eqref{ebeq:ReichShabymodel} using pairwise extremal coefficients. 
To estimate the extremal coefficient function, we consider the process at $n_s$ spatial locations $\bs_1,\ldots,\bs_{n_s}$ and $n_t$ times $t=1,\ldots,n_t$.
The basis functions are fixed over time, but the random effects and errors are independent over time.
That is
\begin{align}\label{ebeq:modelZt}
  Z_t(\bs) = \theta_t(\bs) \epsilon_t(\bs) \quad \mathrm{where} \quad \theta_t(\bs) = \left\{ \sum_{l=1}^L B_{l}(\bs)^{1/\alpha}A_{lt}\right\}^{\alpha},
\end{align}
$A_{lt} \iid$ PS$(\alpha)$, and $\epsilon_t(\bs) \iid$ GEV$(1, \alpha, \alpha)$.
Denote $Y_t(\bs_i) = Y_{it}$, $B_l(\bs_i) = B_{il}$ and $\vartheta(\bs_i,\bs_j) = \vartheta_{ij}$.

In this section we develop an algorithm to estimate the parameter $\alpha$ and the $n_s\times L$ matrix $\bB = \{B_{il}\}$. Our algorithm has the following steps:
\begin{enumerate}
  \item Obtain an initial empirical estimate of the extremal coefficient for each pair of locations, ${\hat \vartheta}_{ij}$.
  \item Spatially smooth the ${\hat \vartheta}_{ij}$ using kernel smoothing to obtain ${\tilde \vartheta}_{ij}$.
  \item Estimate the spatial dependence parameters by minimizing the distance between model-based coefficients $\vartheta_{ij}$ and smoothed empirical coefficients ${\tilde \vartheta}_{ij}$.
\end{enumerate}

The first-stage empirical estimates ${\hat \vartheta}_{ij}$ of the extremal coefficients are obtained from the F-madogram estimator of \citet{Cooley2006}, using the `SpatialExtremes' \citep{Ribatet2015} package of R \citep{Rmanual}; see Appendix~\ref{eba:fmado} for the definition of the estimator. Note that other estimators for ${\hat \vartheta}_{ij}$ could be used, such as the estimators of \citet{Smith1990} or \citet{Schlather2003}. 
Assuming the true extremal coefficient is smooth over space, the initial estimates ${\hat \vartheta}_{ij}$ can be improved by smoothing. For $i\neq j$ let
\begin{align} \label{ebeq:EChat2}
  {\tilde \vartheta}_{ij} = \dfrac{ \sum_{u=1}^{n_s}\sum_{v=1}^{n_s} w_{iu}w_{jv}{\hat \vartheta}_{uv}}
  {\sum_{u=1}^{n_s}\sum_{v=1}^{n_s} w_{iu}w_{jv}},
\end{align}
where $w_{iu} = \exp\{- (\| \bs_i-\bs_u \|/\delta)^2\}$ is the Gaussian kernel function with bandwidth $\delta>0$.
The elements ${\hat \vartheta}_{ii}$ deteriorate the estimator ${\tilde \vartheta}_{ij}$ as ${\hat \vartheta}_{ii}=1$ for all $i$ by construction and because of the possible discontinuity of the extremal coefficient function at the origin. To eliminate this problem we set $w_{ii}=0$.

First, the parameter $\alpha$ is estimated from the estimates of the extremal coefficients of the closest pairs. For pairs $(i,j)$ corresponding to close spatial locations $(\bs_i,\bs_j)$ we known $\vartheta_{ij} \approx 2^{\alpha}$, and therefore we select a set $\mathcal{N}$ of such pairs and use their $\tilde \vartheta_{ij}$ to estimate $\alphahat = \log_2 ( \sum_{\mathcal{N}}{\tilde \vartheta}_{ij}/|\mathcal{S}|)$. For $\hat\alpha$ to be a reasonable estimate of $\alpha$ it is crucial that the data set contains close locations. Second, we estimate $\bB$ given $\alpha=\alphahat$ by minimizing the mean square distance between the estimates ${\tilde \vartheta}_{ij}$ and the model-based extremal coefficients $\vartheta_{ij}$. Similarly to \citet{Smith1990} and \citet{Einmahl.etal:2016b}, we estimate $\hat{\bB}$ as the minimizer of the error
\begin{align} \label{ebeq:Bhat}
\sum_{i<j} \left({\tilde \vartheta}_{ij} - \vartheta_{ij}\right)^2
  =
\sum_{i<j} \left\{{\tilde \vartheta}_{ji} - \sum_{l=1}^L \left(B_{il}^{1/\alphahat} + B_{jl}^{1/\alphahat} \right)^{\alphahat}\right\}^2
\end{align}
under the restrictions that $B_{il}\ge 0$ for all $i$ and $l$, and $\sum_{l=1}^LB_{il}=1$ for all $i$.
Since the minimizer of \eref{ebeq:Bhat} does not have a closed form, we use a gradient descent algorithm to obtain ${\hat \bB}$. 
This algorithm gives estimates of the $B_{il}$ at the $n_s$ data locations, but is easily extended to all $\bs$ for spatial prediction.
The kernel smoothing step ensures that the estimates for $\hat{B}_{il}$ are spatially smooth, and thus interpolation of the $\hat{B}_{il}$ gives spatial functions $\hat{B}_l(\bs)$.

As mentioned in Section~\ref{ebs:model}, the EBFs provide useful exploratory data analysis techniques.
Maps of $\hat{B}_l(\bs)$ show important spatial features in the extremal dependence.
The relative contribution of each EBF can be measured by
\begin{align*}
v_l = \frac{1}{n_s}\sum_{i=1}^{n_s}{\hat B}_{il}.
\end{align*}
Since $\sum_{l=1}^L{\hat B}_{il}=1$ for all $i$, we have $\sum_{l=1}^Lv_l = 1$. Since the EBFs are combined with independent and identically distributed variables $A_1,\ldots,A_L$ in \eqref{ebeq:ReichShabymodel}, the EBFs with largest $v_l$ contribute the most to the model, therefore, EBFs with large $v_l$ are the most important in explaining extreme dependence. 
The order of the basis functions is arbitrary, so we reorder them so that $v_1\ge \cdots \ge v_L$.

The first-stage estimate of the extremal coefficients has two tuning parameters: the kernel bandwidth ($\delta$) for the smoothing step and the number of basis functions ($L$). We use cross-validation on the extremal coefficients to chose an appropriate value for $\delta$. For $L$ we can look for an ``elbow'' in the  error \eqref{ebeq:Bhat} to determine a sensible number of basis functions, or we can use a validation data set (or cross-validation) to assess how $L$ affects predictions from the max-stable model, see \sref{ebs:analysis}.

\section{Bayesian modeling based on empirical basis functions}\label{ebs:MCMC} In this section we describe how to use the EBFs to construct a spatial max-stable model for block maxima. The matrix $\hat{\bB}$, i.e., the EBFs, and the estimate $\hat\alpha$ obtained from the procedure described in Section~\ref{ebs:estimation} provide a max-stable model for the residual extremal spatial dependence using the positive stable model~\eqref{ebeq:modelZt}. Notice that using the EBFs in spatial modeling induces a non-stationary spatial dependence structure. To fully define a max-stable model for block maxima we need to model the marginal GEV distributions in~\eqref{ebeq:GEVmarg}. This can be done using space-time models for the marginal GEV parameters, using covariates and Gaussian process priors, as described, for example, in \citet{Reich2012}. In \sref{ebs:analysis} we model only the residual dependence, so we don't provide more details on marginal modeling in this paper. 

The residual dependence parameters $\hat{\bB}$ and $\hat\alpha$ are fixed and only the positive stable variables $A_{lt}$ and the marginal parameters must be estimated. A nice feature of the hierarchical max-stable model \eqref{ebeq:ReichShabymodel} is that conditional on $\theta(\bs)$, the variables $Y(\bs)$ are conditionally independent, making Bayesian computations for the marginal parameters relatively easy. All parameters can be estimated using standard MCMC methods, a Metropolis-Hastings algorithm with random walk candidate distributions. The main difficulty in fitting the Bayesian hierarchical model concerns the MCMC algorithm for the positive stable variables $A_{lt}$. This is because the positive stable density is challenging to evaluate as it does not have a closed form. One technique to avoid this complication is to incorporate auxiliary random variables, see \citet{Stephenson2009} and \citet{Reich2012}. This approach may lead to poor mixing of the Markov chain of the positive stable variable due to the large number of these auxiliary variables and their dependencies. In \sref{ebs:analysis} we use instead a numerical approximation of the positive stable density, see \aref{eba:gridapprox}.

\section{Simulation study}\label{ebs:sims}

In this section we conduct a simulation study to evaluate the performance of our method for estimating the EBFs.  Each data sets consists of $n_t$ independent replications of the max-stable process at $n_s=100$ spatial locations sampled uniformly on $[1,10]\times [1,10]$.  The data are generated from the \citet{Reich2012} model with unit Fr\'echet marginal distribution, dependence parameter $\alpha$, and $L$ basis functions $B_{il}=B_l(\bs_i)$ defined via \eqref{eq:gkf} with $\rho=2.5$, where the knots $\bk_1,...,\bk_l$ form a rectangular $\sqrt{L}\times \sqrt{L}$ grid spanning $[1,10]\times [1,10]$. 
We generate 100 data sets for each combination of $L\in\{9,25\}$, $\alpha\in\{0.3,0.7\}$, and $n_t\in\{50,200\}$.

For each data set we compute the EBF estimate of $\alpha$ and the EBF estimate of the basis functions $B_{il}$ using the method of \sref{ebs:estimation} with both $L=9$ and $L=25$.  Table \ref{ebtbl:sims} presents the mean and standard deviation of the estimates of $\alpha$ over the 100 data sets.  We also compute the mean squared error for estimating the true extremal coefficients $\vartheta_{ij}$ (determined by the true $\alpha$ and $B_{il}$) averaged over $i$, $j$ and the 100 datasets.  We compare four estimators of the extremal coefficients: (1) the initial estimates from the F-madogram (${\hat \vartheta}_{ij}$), (2) the smoothed estimates (${\tilde \vartheta}_{ij}$), (3) the extremal coefficient corresponding to the EBF estimates of $\alpha$ and $B_{il}$ with $L=9$, (4) the EBF estimates with $L=25$. 

\begin{table}[t!]
	\caption{Simulation study results.  The simulation settings vary by the true number of basis functions, $L$, the spatial-dependence parameter, $\alpha$, and the number of replicates, $n_t$. The results for each of the eight scenarios are summarized using the mean (standard deviation) of the EBF estimates of $\alpha$, and the mean squared error (times 100) of the estimated extremal coefficients $\vartheta_{ij}$ for the initial F-madogram estimate, the kernel smoothed estimate, and the EBF estimates assuming $L=9$ or $L=25$ basis functions.}
	\label{ebtbl:sims}
	\centering
	\begin{tabular}{ccc|c|cccc}
		\multicolumn{3}{c}{Settings} & & \multicolumn{4}{c}{Extremal coefficients} \\
		$L$ & $\alpha$ & $n_t$ & ${\hat \alpha}$ & Initial 
		& Smoothed & EBF, $L=9$ &EBF, $L=25$\\\hline
 9 &  0.3 &  50 & 0.31 (0.02) & 1.08 & 0.87 & 0.84 & 0.88\\
 9 &  0.3 & 200 & 0.31 (0.01) & 0.28 & 0.28 & 0.27 & 0.31\\
 9 &  0.7 &  50 & 0.70 (0.04) & 1.19 & 0.49 & 0.50 & 0.51\\
 9 &  0.7 & 200 & 0.70 (0.02) & 0.31 & 0.14 & 0.13 & 0.15\\
25 &  0.3 &  50 & 0.32 (0.02) & 1.12 & 0.89 & 2.08 & 0.95\\
25 &  0.3 & 200 & 0.32 (0.01) & 0.27 & 0.29 & 1.56 & 0.38\\
25 &  0.7 &  50 & 0.70 (0.03) & 1.12 & 0.55 & 0.84 & 0.60\\
25 &  0.7 & 200 & 0.70 (0.01) & 0.27 & 0.16 & 0.46 & 0.20
	\end{tabular}
\end{table}

The EBF method for estimating the spatial dependence parameter $\alpha$ works well in all cases.  Smoothing the initial F-madogram estimate leads to an improvement except for the scenario with strong spatial dependence ($\alpha=0.3$) and large sample size ($n_t=200$) where the initial estimate is sufficient. The EBF method with the correct number of basis functions has similar MSE to the smoothed estimate which shows that the method is able to efficiently represent spatial dependence with a small number of basis functions.  As expected, MSE increases slightly when too many basis functions are included (the first four rows of Table \ref{ebtbl:sims}), and MSE increases substantially when too few basis functions are included (the final four rows of Table \ref{ebtbl:sims}).  This suggests that the user should error on the side of including too many basis functions.  

\section{Analysis of annual maximum precipitations}\label{ebs:analysis}  

We illustrate the EBF method with an analysis of annual maxima of precipitation in the eastern U.S.
We use the same dataset as \citet{Reich2012} and we compare our EBF approach to their original Gaussian kernel function (GKF) approach. 
The data are climate model output from the North American Regional Climate Change Assessment Program (NARCCAP).
This dataset consists of $n_s = 697$ grid cells at a 50~km resolution in the eastern US, and includes historical data (1969--2000) as well as future conditions (2039--2070). 

We focus on the analysis of the residual dependence in the data and in particular in comparing our EBF method with the original GKF method in terms of dependence modeling. In order to focus on the residual dependence we transformed the data to common unit Fr\'echet marginals in a first step, and then performed all subsequent analysis on the unit Fr\'echet scale assuming the marginals are fixed and known exactly. 
This was done to isolate differences in residual dependence modeling, which is the focus of the current paper. 
Initial data analysis on the transformed data suggests that the dependence is different between current and future periods. The residual extreme dependence appear stronger in the future period than in the current period. Thus we decided to analyze the two periods separately.

We compare the proposed EBF method described in Sections 2-4 with the Gaussian kernel functions (GKFs) model of \citet{Reich2012}. \citet{Reich2012} fit a model that uses standardized Gaussian kernel functions (GKFs), see~\eqref{eq:gkf}, centered on a set of $\bk_1, \ldots, \bk_L$ spatial knots for the spatial basis functions $B_l(\bs)$. We denote by $\rho$ the bandwidth of these Gaussian kernels. 
To compare the EBF and GKF methods we apply two modifications to the original GKF approach. First, \citet{Reich2012} used a large number of GKF, but here we use a relatively small number $L$ of basis functions for both EBF and GKF approaches. The spatial knots $\bk_1, \ldots, \bk_L$ are selected using the `cover.design' function in the `fields' \citep{fields2015} package of R, with default settings; the method selects the centroid locations which minimize a geometric space-filling criterion. Second, \citet{Reich2012} estimated $\rho$ and $\alpha$ in the MCMC algorithm, but here we use the procedure described in Section~\ref{ebs:estimation} to estimate the dependence parameters for both methods and then treat these parameters as fixed in the MCMC algorithm. In particular, $\hat\alpha$ is the same for the EBF and GKF methods and does not depend on $L$.

\subsection{Comparison of EBF and GKF methods}\label{ebs:precip}
The analysis of the transformed data (which we called the residuals) is performed in two steps: first we estimate $\alpha$, the EBFs, and the GKFs bandwidth $\rho$ as described in \sref{ebs:estimation}, and second we fit Bayesian hierarchical models based on these basis functions as described in \sref{ebs:MCMC}, with marginals that are fixed to unit Fr\'echet; i.e., the MCMC algorithm only estimates the latent variables $A_{lt}$. We apply these methods to the two periods separately.

\begin{figure}[t!]
  \centering
  \includegraphics[width=0.9\linewidth]{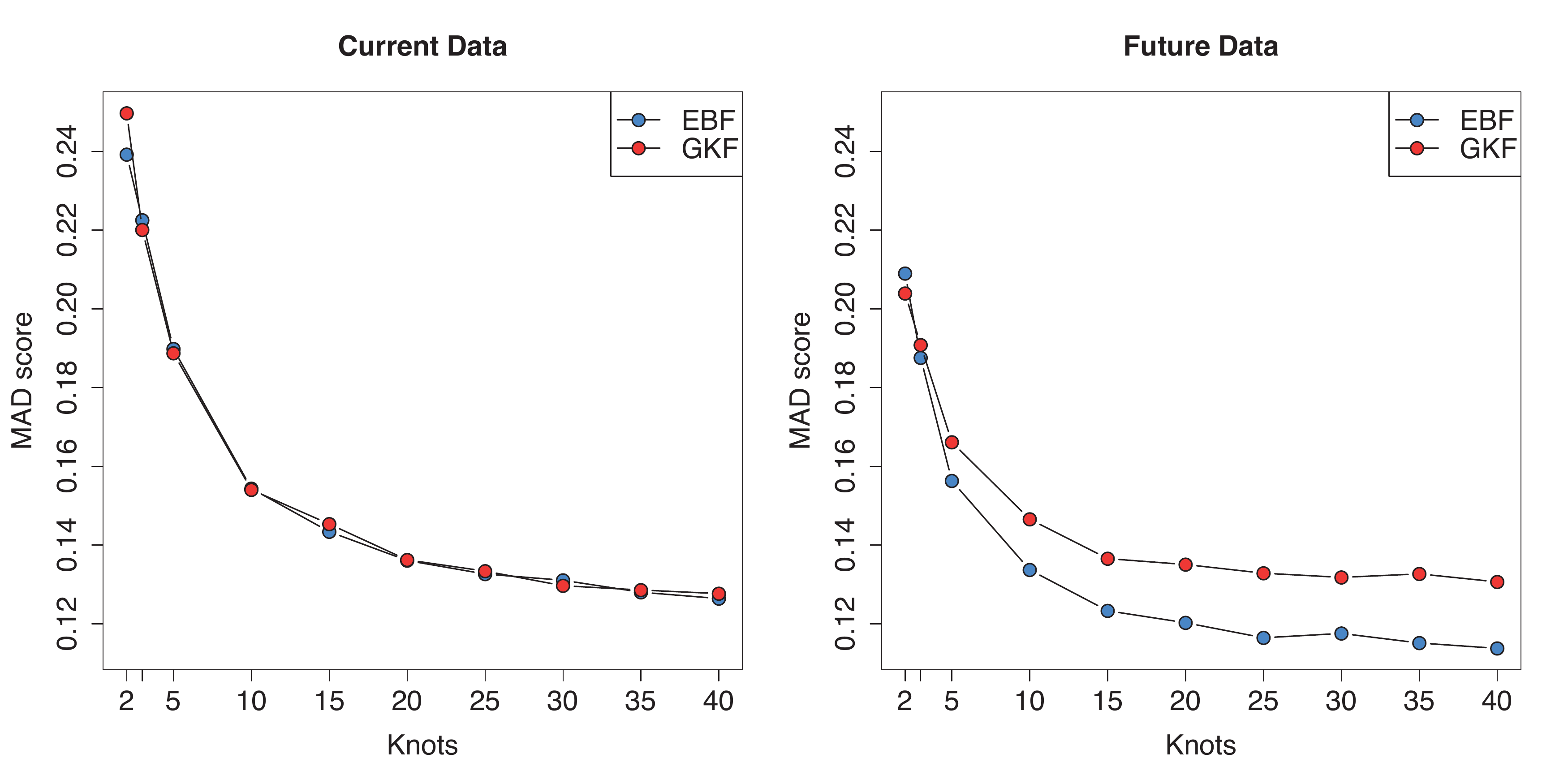}
  \caption{Mean absolute deviation (MAD) scores by the number of basis functions ($L$) estimated using cross-validation. Current and future data are modeled and evaluated separately.}
  \label{fig:precip-scores}
\end{figure}

We use 5-fold cross-validation to assess the predictive performance of the EBF and GKF models for current and future data and for different values of $L$. The dataset is randomly split into five (roughly-equal size) parts accross sites and years. The models are then trained onto observations of four of the five parts and the remaining observations are used to evaluate the predictions. Training the models imply estimating $\alpha$, the EBF and the GKF and then using the MCMC to predict data on the test set. To assess the predictions for the test set we use the mean absolute deviation (MAD). The two periods are treated separately in terms of fit and evaluation. The scores for the current and future precipitation data analysis are given in Figure~\ref{fig:precip-scores}.

We observe some variation in the scores across the number of basis functions $L$.  On the current data, EBF and GKF method give similar results (scores are only significantly different for $L=2$). The MAD score reduces sharply from $L=2$ to $L=10$, and only slightly after $L=30$.  On the future data, EBF gives lower MAD score than GKF for all $L$. The MAD score reduces sharply from $L=2$ to $L=10$, and seems to stabilize around $L=15$. These results suggest that 10 EBFs might be sufficient to adequately model spatial dependence in current and future precipitation data.

\subsection{Extremal dependence in current and future precipitations} \label{ebs:results-precip}
\begin{figure}[t!]
  \centering
  \includegraphics[width=0.9\linewidth]{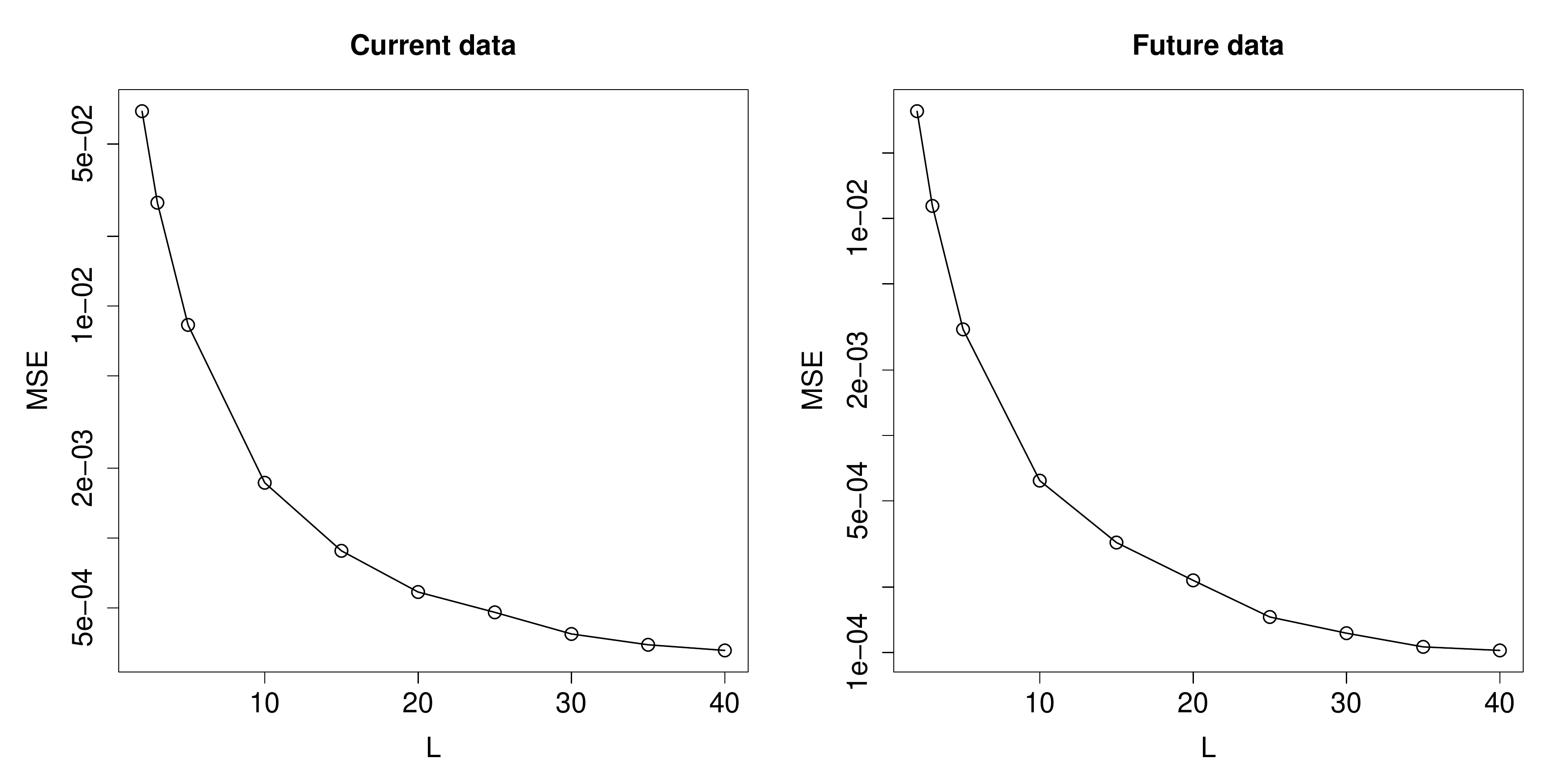}
  \caption{Mean squared error (proportional to~\eqref{ebeq:Bhat}) of EBFs by the number of basis functions ($L$). Note the $y$-axis is on log scale.}
  \label{fig:EBFerr}
\end{figure}

In this section we discuss the analysis of all current and future data; again the two periods are analyzed separately. The cross-validation results suggest using $L=10$ for both current and future data. As an alternative method to choose $L$ we estimated the EBFs for various values of $L$ and we computed the error \eqref{ebeq:Bhat} between the smoothed empirical extremal coefficients and the EBF based extremal coefficients, see \fref{fig:EBFerr}. This plot also suggests that $L=10$ is a reasonable choice for both periods, without the need of extensive modeling fitting.

\begin{figure}[t!]
  \centering
  \includegraphics[width=1\linewidth]{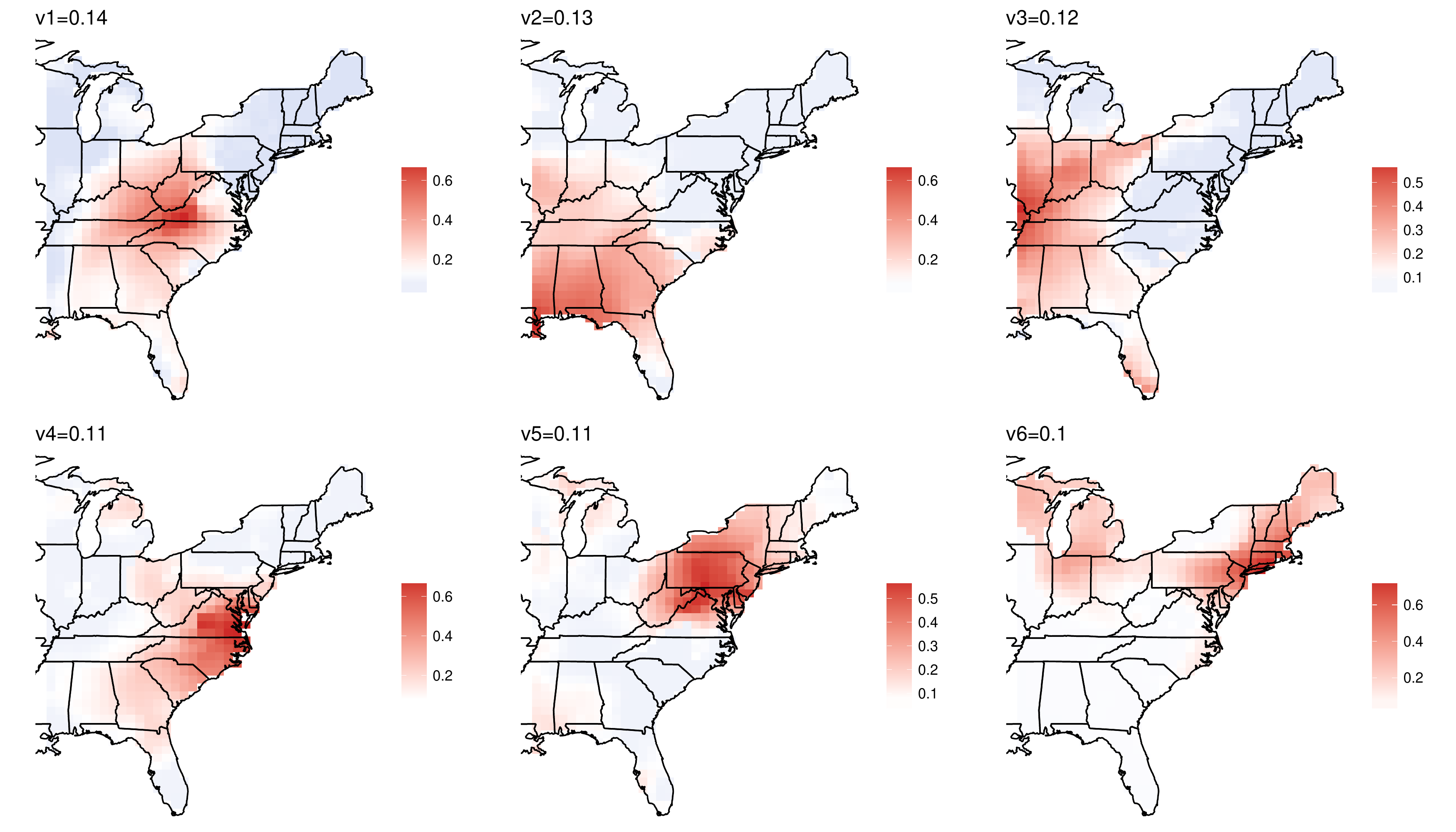}\\
  \caption{First six EBFs along with their contributions $v_1, \ldots, v_{6}$, for $L=10$ and for the current precipitation data. Note the different scales in each figure.}
  \label{ebfig:EBFs10current}
\medskip
  \includegraphics[width=1\linewidth]{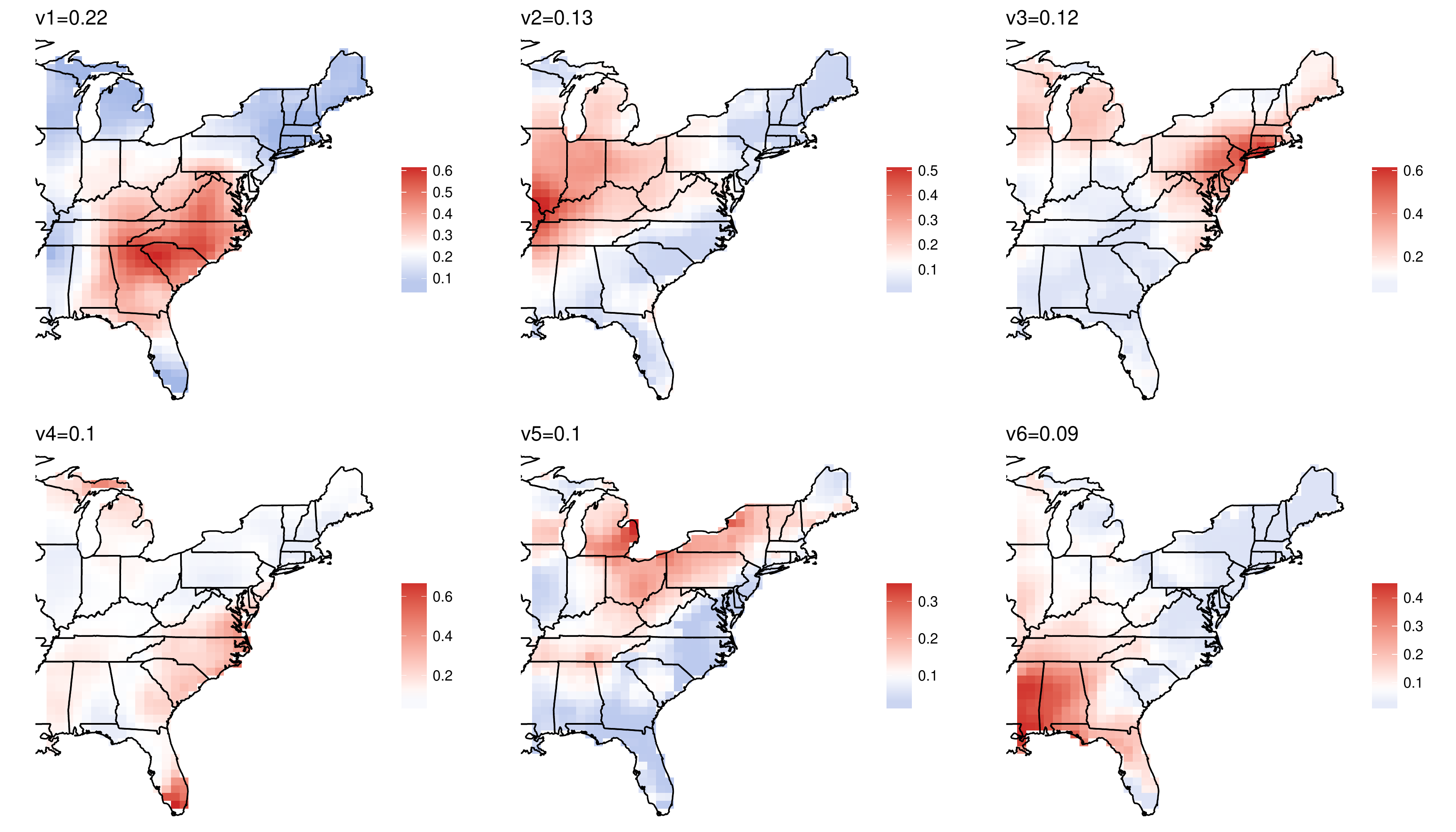}\\
  \caption{First six EBFs along with their contributions $v_1, \ldots, v_{6}$, for $L=10$ and for the future precipitation data. Note the different scales in each figure.}
  \label{ebfig:EBFs10future}

\end{figure}

We discuss the EBFs for $L=10$ for current and future data. First, for current data $\hat\alpha=0.24$, and for future data $\hat\alpha=0.21$, showing the residuals are smoother for the future period. The first six EBFs (out of ten) along with their contributions $v_1, \ldots, v_{6}$ are given in \fref{ebfig:EBFs10current} for current data, and in \fref{ebfig:EBFs10future} for future data. These six EBFs show the main features needed to reconstruct extremal dependence in the annual maximum precipitations in this region. Most of the EBFs capture local features. For example, the first EBF for the current regime explains extreme precipitation in the Appalachian Mountains, while the second separates the south from the rest of the spatial domain.
 
\begin{figure}[t!]
  \centering
  \includegraphics[width=0.58\linewidth]{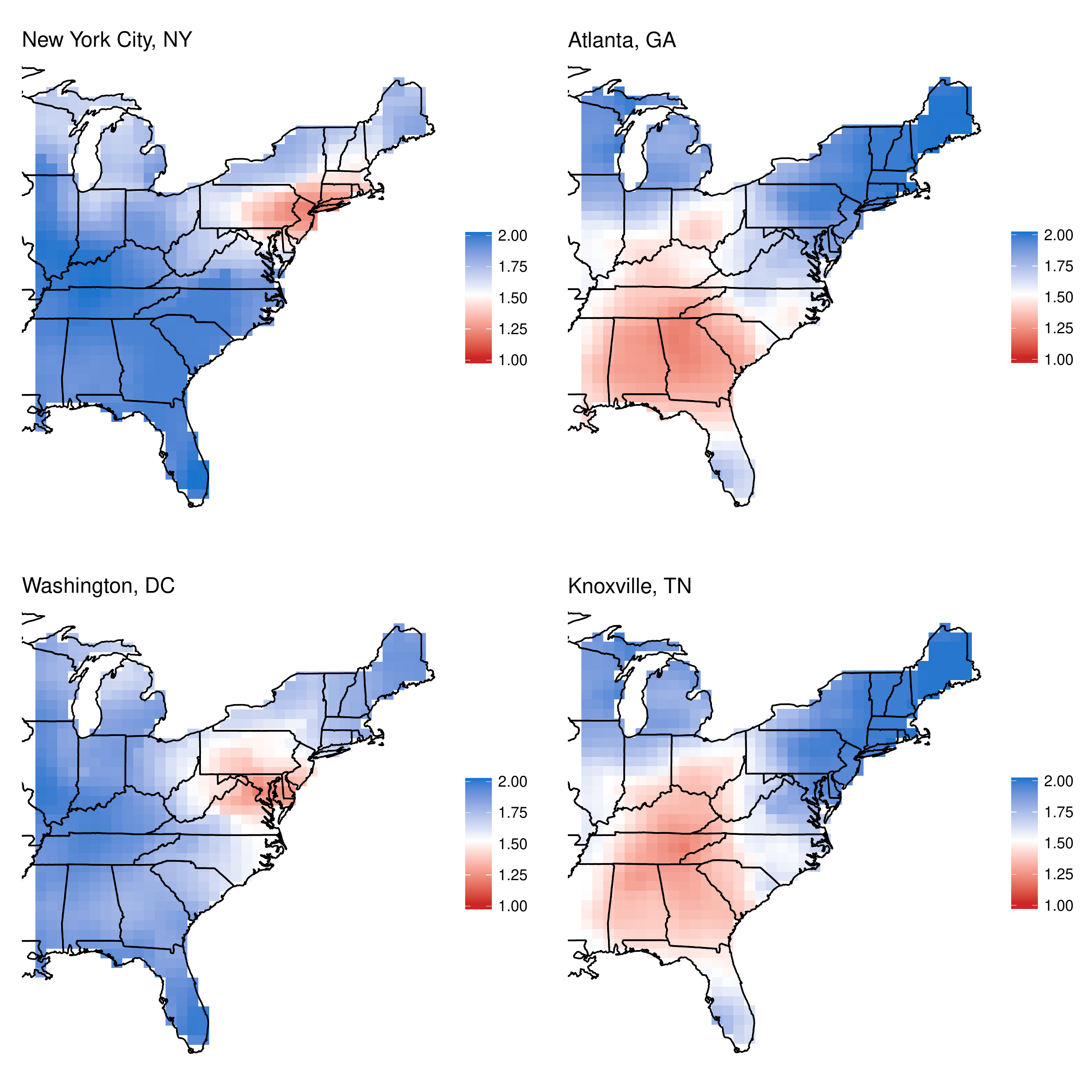}
  \caption{Estimated pairwise extremal coefficients for current data (using $L = 10$) for New York City, NY; Atlanta, GA; Washington, DC; and Knoxville, TN.}
  \label{ebfig:ECmaps-current}
  \medskip
  \includegraphics[width=0.58\linewidth]{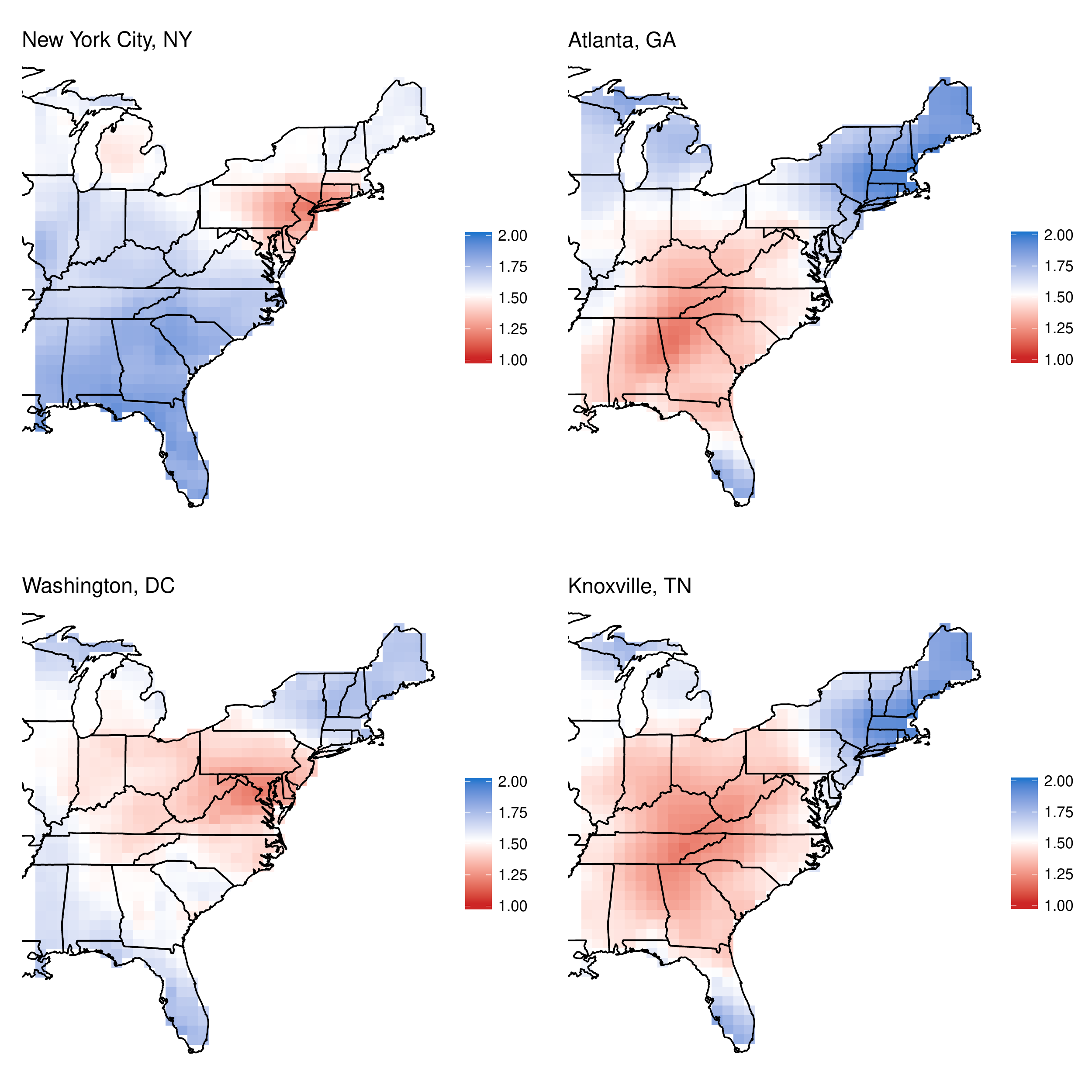}
  \caption{Estimated pairwise extremal coefficients for future data (using $L = 10$) for New York City, NY; Atlanta, GA; Washington, DC; and Knoxville, TN.}
  \label{ebfig:ECmaps-future}
\end{figure}

Figures~\ref{ebfig:ECmaps-current} and \ref{ebfig:ECmaps-future} show extremal coefficient maps $\{\vartheta(\bs_i,\bs)\}_{\bs\in\calS}$ for some sites $\bs_i$. The estimates of spatial dependence clearly exhibit non-stationary.  For example, in the current climate the range of spatial dependence is much smaller around New York City and Washington, DC than it is around Atlanta and Knoxville.  Also, around all four of these cities the range of spatial dependence appears to increase in the future climate.

\section{Discussion}\label{ebs:con}

In this paper we have proposed a new empirical basis function approach for a data-driven low-rank max-stable process.
The EBFs provide researchers with an exploratory data analysis tool to explore spatial extremal dependence.
The EBFs can also be used as inputs to a Bayesian model for inference and prediction over space and time.

The results from the data analysis suggest that in the presence of strong spatial dependence, as with the considered future precipitation data, the EBFs show an improvement in prediction accuracy over using the  GKF approach. Compared to the original approach of \citet{Reich2012} that uses as many kernels as the number of sites, or to \citet{Shaby.Reich:2012} and \citet{Stephenson2015} that uses a large number of kernels, centered on a grid or at a subsample of the observation locations, the computation of the Bayesian model based on the EBFs with relatively few basis functions is much faster. On average, it takes 2~seconds for 10000 updates of each latent variable $A_{lt}$, such that the running time for fitting our EBF dependence model (with fixed marginals) is about $2 \times L \times n_t$~seconds for 10000 iterations of the MCMC algorithm. Cross-validation can be used to select a number of basis functions $L$ such that the predictive accuracy of the Bayesian model using a small number of EBFs is comparable to a more complex Bayesian model using a large number of GKFs.

We have used the EBF for exploratory analysis and Bayesian inference. Another possibility is to use the methods to reduce the data under consideration from the actual responses to loadings $A_{lt}$.
Given the EBFs, one could obtain estimates of the $A_{lt}$ for each time point.  Time series of the estimated $A_{lt}$ may be used as a fast and simple method to study large-scale spatiotemporal trends.

The EBF approach introduced here can be easily extended to explore and model extremal dependence in threshold exceedances. Max-stable processes are sensible models for exceedances over large thresholds and empirical extremal coefficients for threshold exceedances can be estimated from censored versions of the likelihood estimator of \citet{Schlather2003}, for example. From these empirical extremal coefficients, EBFs can be estimated just as described in \sref{ebs:estimation}. The positive stable max-stable model based on these EBFs can then be fitted to threshold exceedances using a censored approach as in \citet{Reich2014} and \citet{Morris2016}.

\section*{Acknowledgements}
The authors acknowledge Dan Cooley for his helpful suggestions on the manuscript.
The authors' work was partially supported by grants from the Department of the Interior (14-1-04-9), National Institutes of Health (R21ES022795-01A1), the US Environmental Protection Agency (R835228), the National Science Foundation (1107046). The calculations have been performed using the facilities of the Scientific IT and Application Support Center of EPFL. 

\appendix
\section{Appendix}
\subsection{Grid approximation to positive stable density} \label{eba:gridapprox}
The positive stable density does not have a closed form. From \citet[Section 2]{Stephenson2009} the density of PS($\alpha$) can be expressed as
\begin{align}\label{eq:intPS}
f(x) = \int_0^1 h(x, y) \dd y,
\end{align}
where
\begin{align*}
h(x, y) = \frac{\alpha}{1 - \alpha} \left( \frac{1}{x} \right)^{1 / 1 - \alpha} c(\pi y) \exp \left\{ - c(\pi y) \left(\frac{1}{x}\right)^{\alpha / (1 - \alpha)} \right\},
\end{align*}
with
\begin{align*}
c(\psi) = \left\{\frac{\sin(\alpha \psi)}{\sin(\psi)}\right\}^{1 / (1 - \alpha)} \frac{\sin\{(1 - \alpha) \psi\}}{\sin(\alpha \psi)}.
\end{align*}
\citet{Stephenson2009} presents an auxiliary variable technique to deal with the integral in the density function, but we opt to numerically evaluate the integral \eqref{eq:intPS} because it is only one-dimensional. 
We use the midpoint rule with 50 evenly spaced quantiles of a Beta$(0.5, 0.5)$ distribution as the midpoints.

\subsection{F-madogram estimator} \label{eba:fmado}
For two locations $\bs_i,\bs_j\in\calS$, the F-madogram of \citet{Cooley2006}, here defined for a non-stationary max-stable process $Y$, is
$$
\nu^{F}(\bs_i,\bs_j) = \dfrac12 {\rm E}| F_{\bs_i}\{Y(\bs_i)\} - F_{\bs_j}\{Y(\bs_j)\} |
$$
where $F_{\bs_i}$ denote the distribution of $Y(\bs_i)$. \citet{Cooley2006} showed the extremal coefficient is related to the F-madogram by $\vartheta(\bs_1,\bs_2)=\{1+2\nu^{F}(\bs_i,\bs_j)\}/\{1-2\nu^{F}(\bs_i,\bs_j)\}$.

Assume we observe $n_t$ realizations $y_{i1},\ldots,y_{in_t}$ of $Y(\bs_i)$ at some locations $\bs_i\in\calS$. We define the nonparametric estimator $\hat F_{\bs_i}(y_{it})={\rm rank}(y_{it})/(n_t+1)$, where ${\rm rank}(y_{it})$ denotes the rank of the value $y_{it}$ among $(y_{i1},\ldots,y_{in_t})$. We estimate the F-madogram by
$$
\hat\nu^{F}(\bs_i,\bs_j) = \dfrac1{2n_t}  \sum_{t=1}^{n_t} | \hat F_{\bs_i}(y_{it}) - \hat F_{\bs_i}(y_{jt}) | = \dfrac1{2n_t}  \sum_{t=1}^{n_t} | {\rm rank}(y_{it})/(n_t+1) - {\rm rank}(y_{jt})/(n_t+1) |.
$$
The extremal coefficient is then estimated by $\hat\vartheta(\bs_1,\bs_2)=\{1+2\hat\nu^{F}(\bs_i,\bs_j)\}/\{1-2\hat\nu^{F}(\bs_i,\bs_j)\}$.


\begin{thebibliography}{}

\bibitem[Bernard et~al., 2013]{Bernard2013}
Bernard, E., Naveau, P., Vrac, M., and Mestre, O. (2013).
\newblock {Clustering of Maxima: Spatial Dependencies among Heavy Rainfall in
  France}.
\newblock {\em Journal of Climate}, 26(20):7929--7937.

\bibitem[Coles, 2001]{Coles2001}
Coles, S. (2001).
\newblock {\em {An Introduction to Statistical Modeling of Extreme Values}}.
\newblock Lecture Notes in Control and Information Sciences. Springer, London.

\bibitem[Cooley et~al., 2006]{Cooley2006}
Cooley, D., Naveau, P., and Poncet, P. (2006).
\newblock {Variograms for spatial max-stable random fields}.
\newblock In Bertail, P., Soulier, P., and Doukhan, P., editors, {\em
  Dependence in Probability and Statistics}, volume 187 of {\em Lecture Notes
  in Statistics}, chapter Variograms, pages 373--390. Springer New York, New
  York, NY.

\bibitem[Davison et~al., 2013]{Davison2013}
Davison, A.~C., Huser, R., and Thibaud, E. (2013).
\newblock {Geostatistics of Dependent and Asymptotically Independent Extremes}.
\newblock {\em Mathematical Geosciences}, 45(5):511--529.

\bibitem[Davison et~al., 2012]{Davison2012}
Davison, A.~C., Padoan, S.~A., and Ribatet, M. (2012).
\newblock {Statistical modeling of spatial extremes}.
\newblock {\em Statistical Science}, 27(2):161--186.

\bibitem[de~Haan and Ferreira, 2006]{deHaan2006}
de~Haan, L. and Ferreira, A. (2006).
\newblock {\em {Extreme Value Theory: An Introduction}}.
\newblock Springer Series in Operations Research and Financial Engineering.
  Springer.

\bibitem[Dey and Yan, 2015]{Dey.Yan:2015}
Dey, D.~K. and Yan, J. (2015).
\newblock {\em {Extreme Value Modeling and Risk Analysis: Methods and
  Applications}}.
\newblock Chapman and Hall/CRC.

\bibitem[Einmahl et~al., 2016]{Einmahl.etal:2016b}
Einmahl, J. H.~J., Kiriliouk, A., and Segers, J. (2016).
\newblock A continuous updating weighted least squares estimator of tail
  dependence in high dimensions.
\newblock arXiv:1601.04826.

\bibitem[Engelke et~al., 2015]{Engelke2015}
Engelke, S., Malinowski, A., Kabluchko, Z., and Schlather, M. (2015).
\newblock {Estimation of H{\"{u}}sler-Reiss distributions and Brown-Resnick
  processes}.
\newblock {\em Journal of the Royal Statistical Society. Series B: Statistical
  Methodology}, 77:239--265.

\bibitem[Everitt and Hothorn, 2008]{Everitt2008}
Everitt, B. and Hothorn, T. (2008).
\newblock {Principal components analysis}.
\newblock In {\em An Introduction to Applied Multivariate Analysis with R},
  pages 21--54. Springer New York, New York, NY.

\bibitem[Foug{\`e}res et~al., 2013]{Fougeres.etal:2013}
Foug{\`e}res, A.-L., Mercadier, C., and Nolan, J.~P. (2013).
\newblock Dense classes of multivariate extreme value distributions.
\newblock {\em Journal of Multivariate Analysis}, 116:109--129.

\bibitem[Foug\`eres et~al., 2009]{Fougeres.etal:2009}
Foug\`eres, A.-L., Nolan, J.~P., and Rootz\'en, H. (2009).
\newblock {Models for Dependent Extremes Using Stable Mixtures}.
\newblock {\em Scandinavian Journal of Statistics}, 36(1):42--59.

\bibitem[Hannachi et~al., 2007]{Toggweiler2001}
Hannachi, A., Jolliffe, I.~T., and Stephenson, D.~B. (2007).
\newblock {Empirical orthogonal functions and related techniques in atmospheric
  science: A review}.
\newblock {\em International Journal of Climatology}, 27(9):1119--1152.

\bibitem[Huser and Davison, 2014]{Huser2014}
Huser, R. and Davison, A.~C. (2014).
\newblock {Space-time modelling of extreme events}.
\newblock {\em Journal of the Royal Statistical Society: Series B (Statistical
  Methodology)}, 76(2):439--461.

\bibitem[Kabluchko et~al., 2009]{Kabluchko2009}
Kabluchko, Z., Schlather, M., and de~Haan, L. (2009).
\newblock {Stationary max-stable fields associated to negative definite
  functions}.
\newblock {\em Annals of Probability}, 37(5):2042--2065.

\bibitem[Lee and Seung, 1999]{Lee1999}
Lee, D.~D. and Seung, S.~H. (1999).
\newblock Learning the parts of objects by non-negative matrix factorizations.
\newblock {\em Nature}, 401:788 -- 791.

\bibitem[Mairal et~al., 2014]{Mairal2014}
Mairal, J., Bach, F., and Ponce, J. (2014).
\newblock Sparse modeling for image and vision processing.
\newblock {\em Foundations and Trends in Computer Graphics and Vision}, 8:85 --
  283.

\bibitem[Morris et~al., 2017]{Morris2016}
Morris, S.~A., Reich, B.~J., Thibaud, E., and Cooley, D. (2017).
\newblock A space-time skew-\emph{t} model for threshold exceedances.
\newblock {\em Biometrics}, 73(3):749--758.

\bibitem[Nychka et~al., 2015]{fields2015}
Nychka, D., Furrer, R., and Sain, S. (2015).
\newblock {\em fields: Tools for Spatial Data}.
\newblock R package version 8.2-1.

\bibitem[Padoan et~al., 2010]{Padoan2010}
Padoan, S.~A., Ribatet, M., and Sisson, S.~A. (2010).
\newblock {Likelihood-based inference for max-stable processes}.
\newblock {\em Journal of the American Statistical Association},
  105(489):263--277.

\bibitem[{R Core Team}, 2016]{Rmanual}
{R Core Team} (2016).
\newblock {\em R: A Language and Environment for Statistical Computing}.
\newblock R Foundation for Statistical Computing, Vienna, Austria.

\bibitem[Reich and Shaby, 2012]{Reich2012}
Reich, B.~J. and Shaby, B.~A. (2012).
\newblock {A hierarchical max-stable spatial model for extreme precipitation}.
\newblock {\em The Annals of Applied Statistics}, 6(4):1430--1451.

\bibitem[Reich et~al., 2014]{Reich2014}
Reich, B.~J., Shaby, B.~A., and Cooley, D. (2014).
\newblock {A Hierarchical Model for Serially-Dependent Extremes: A Study of
  Heat Waves in the Western US}.
\newblock {\em Journal of Agricultural, Biological, and Environmental
  Statistics}, 19(1):119--135.

\bibitem[Ribatet, 2015]{Ribatet2015}
Ribatet, M. (2015).
\newblock {\em SpatialExtremes: Modelling Spatial Extremes}.
\newblock R package version 2.0-2.

\bibitem[Schlather, 2002]{Schlather2002}
Schlather, M. (2002).
\newblock {Models for Stationary Max-Stable Random Fields}.
\newblock {\em Extremes}, 5(1):33--44.

\bibitem[Schlather and Tawn, 2003]{Schlather2003}
Schlather, M. and Tawn, J.~A. (2003).
\newblock {A dependence measure for multivariate and spatial extreme values:
  Properties and inference}.
\newblock {\em Biometrika}, 90(1):139--156.

\bibitem[Shaby and Reich, 2012]{Shaby.Reich:2012}
Shaby, B.~A. and Reich, B.~J. (2012).
\newblock {Bayesian spatial extreme value analysis to assess the changing risk
  of concurrent high temperatures across large portions of European cropland}.
\newblock {\em Environmetrics}, 23(8):638--648.

\bibitem[Smith, 1990]{Smith1990}
Smith, R.~L. (1990).
\newblock {Max-stable processes and spatial extremes}.
\newblock Unpublished manuscript, University of Surrey, Guildford GU2 5XH,
  England. http://www.stat.unc.edu/postscript/rs/spatex.pdf.

\bibitem[Stephenson, 2009]{Stephenson2009}
Stephenson, A.~G. (2009).
\newblock {High-dimensional parametric modelling of multivariate extreme
  events}.
\newblock {\em Australian {\&} New Zealand Journal of Statistics},
  51(1):77--88.

\bibitem[Stephenson et~al., 2015]{Stephenson2015}
Stephenson, A.~G., Shaby, B.~a., Reich, B.~J., and Sullivan, A.~L. (2015).
\newblock {Estimating Spatially Varying Severity Thresholds of a Forest Fire
  Danger Rating System Using Max-Stable Extreme-Event Modeling}.
\newblock {\em Journal of Applied Meteorology and Climatology}, 54(2):395--407.

\bibitem[Thibaud et~al., 2016]{Thibaud2015}
Thibaud, E., Aalto, J., Cooley, D.~S., Davison, A.~C., and Heikkinen, J.
  (2016).
\newblock {Bayesian inference for the Brown--Resnick process, with an
  application to extreme low temperatures}.
\newblock {\em Annals of Applied Statistics}, 10(4):2303--2324.

\bibitem[Thibaud et~al., 2013]{Thibaud2013}
Thibaud, E., Mutzner, R., and Davison, A.~C. (2013).
\newblock {Threshold modeling of extreme spatial rainfall}.
\newblock {\em Water Resources Research}, 49(8):4633--4644.

\bibitem[Thibaud and Opitz, 2015]{Thibaud2013a}
Thibaud, E. and Opitz, T. (2015).
\newblock {Efficient inference and simulation for elliptical Pareto processes}.
\newblock {\em Biometrika}, 102(4):855--870.

\bibitem[Wadsworth and Tawn, 2014]{Wadsworth2014}
Wadsworth, J.~L. and Tawn, J.~A. (2014).
\newblock {Efficient inference for spatial extreme value processes associated
  to log-Gaussian random functions}.
\newblock {\em Biometrika}, 101(1):1--15.

\bibitem[Wang and Stoev, 2011]{Wang2011}
Wang, Y. and Stoev, S.~A. (2011).
\newblock {Conditional sampling for spectrally discrete max-stable random
  fields}.
\newblock {\em Advances in Applied Probability}, 43(2):461--483.

\end{thebibliography}

\end{document}